\documentclass[aps,prd,preprintnumbers,groupedaddress,nofootinbib,showpacs]{revtex4}
\usepackage{graphicx}
\usepackage{latexsym}
\usepackage{amsfonts}
\usepackage{amssymb}
\usepackage{amsmath}
\usepackage{slashed}
\usepackage{array}
\usepackage{feynmp}
\usepackage[hypertex]{hyperref}

\def\lsim{\mathrel{\raise.3ex\hbox{$<$\kern-.75em\lower1ex\hbox{$\sim$}}}}
\def\gsim{\mathrel{\raise.3ex\hbox{$>$\kern-.75em\lower1ex\hbox{$\sim$}}}}

\pdfoutput=1

\usepackage{xcolor}
\definecolor{red}{rgb}{1.0, 0, 0}

\begin{document}


\title{Light $Z'$ Bosons at the Tevatron}
\author{Matthew R.~Buckley,$^{1}$, Dan Hooper$^{1,2}$, Joachim Kopp$^3$, and Ethan Neil$^3$}
\affiliation{$^1$Center for Particle Astrophysics, Fermi National Accelerator Laboratory, Batavia, IL 60510}
\affiliation{$^2$Department of Astronomy and Astrophysics, University of Chicago, Chicago, IL 60637}
\affiliation{$^3$Theoretical Physics Department, Fermi National Accelerator Laboratory, Batavia, IL 60510}
\preprint{}
\date{\today}

\begin{abstract}
New gauge bosons with Standard Model-like couplings to leptons are constrained by collider searches to be heavier than approximately $\sim$1~TeV. A $Z'$ boson with suppressed couplings to leptons, however, could be much lighter and possess substantial couplings to Standard Model quarks. In this article, we consider a new leptophobic $Z'$ gauge boson as a simple and well motivated extension of the Standard Model, and discuss several of its possible signatures at the Tevatron. We find that three of the recent anomalies reported from the Tevatron---in particular the top-quark forward-backward asymmetry and excesses in the $3b$ and $W + 2$~jets final states---could be explained by a new $Z'$ with a mass of approximately 150 GeV, relatively large couplings to quarks, and suppressed couplings to electrons and muons. Moreover, we find that such a particle could also mediate the interactions of dark matter, leading to potentially interesting implications for direct detection experiments. 

\end{abstract}

\pacs{14.70.Pw,14.80.-j,95.35.+d~~~~~~~~~~~~~~FERMILAB-PUB-11-154-A-T}

\maketitle

\section{Introduction \label{sec:intro}}

Of the many high energy extensions of the Standard Model to have been proposed, a new abelian $U(1)'$ gauge group is among the simplest and best motivated. For example, it has been long realized that the matter content of the Standard Model is anomaly free under a gauged $U(1)_{B-L}$. More generally, new gauged $U(1)$ groups appear in many Grand Unified Theories (GUTs), including those based on the gauge groups $SO(10)$ and $E_6$~\cite{London:1986dk,Hewett:1988xc}, and within many other often studied extensions of the Standard Model~\cite{Cvetic:1997ky,Langacker:1999hs,ArkaniHamed:2001nc,ArkaniHamed:2002qx,Han:2003wu,Anastasopoulos:2008jt,FileviezPerez:2010gw}. Assuming a non-negligible coupling strength, any new $U(1)'$ must undergo spontaneous symmetry breaking, allowing the resulting gauge boson -- the $Z'$ -- to be massive.\footnote{For a comprehensive review of $Z'$ phenomenology, see Ref.~\cite{Langacker:2008yv}.}

In most models that introduce a new $U(1)'$ gauge symmetry, the corresponding $Z'$ boson couples to leptons and quarks with similar strengths. In part, this characteristic is helpful in arranging the cancellation of anomalies. Barring the addition of new fermionic particle content beyond the Standard Model, the cancellation of triangle anomalies requires contributions from both leptons and quarks~\cite{Batra:2005rh,Appelquist:2002mw}. Through the introduction of new fermions to aid in anomaly cancellation, however, it is possible to construct $Z'$ models with considerably differing couplings to the various quarks and leptons of the Standard Model.


From a phenomenological perspective, the leptonic couplings of a $Z'$ are particularly important. First of all, $e^+e^-$ colliders such as LEP~II provide the cleanest environment in which to search for pair production of Standard Model particles through the $s$-channel exchange of an off-shell heavy particle such as a $Z'$. Clearly, such experiments require leptonic couplings for the production of the $Z'$, even if the final state is hadronic. Second, although hadron colliders can in principle probe a $Z'$ which only couples to quarks (for instance through the process $q\bar{q} \to (Z')^* \to q'\bar{q}'$ leading to a dijet final state), the considerable QCD backgrounds make the corresponding signal extraction difficult. Therefore, even at hadronic machines, the most stringent $Z'$ constraints derive from leptonic channels. 

For a $Z'$ that couples to both leptons and quarks with strengths similar to those of the Standard Model $Z$, results from LEP~II and the Tevatron constrain its mass to be on the order of 1~TeV or higher. A $Z'$ with somewhat reduced couplings to electrons and muons, however, could easily evade such constraints, even with a mass as light as $\sim$$100-200$~GeV and significant couplings to quarks. While such a leptophobic $Z'$ (or hadrophilic $Z'$, if one prefers to take a more glass half-full outlook) is not generically predicted by models of new physics, such a particle can arise naturally in certain contexts; see Ref.~\cite{delAguila:1986iw} for example.  Furthermore, regardless of such theoretical considerations, the possibility of a leptophobic $Z'$ boson remains viable from a phenomenological perspective.  We will discuss the constraints on such a particle in more detail in Sec.~\ref{sec:bounds}.

Although new gauge bosons with masses well below a TeV and with substantial couplings to quarks are not currently excluded by experimental results, evidence of their existence could potentially appear in a variety of channels at the Tevatron or Large Hadron Collider (LHC). In this paper, we discuss a number of anomalous signals reported from the Tevatron which could be the result of a relatively light and somewhat leptophobic $Z'$. In particular, the CDF collaboration has recently reported the observation of a 3.2$\sigma$ excess in the distribution of events with a leptonically decaying $W^{\pm}$ and a pair of jets \cite{cdfthesis, Aaltonen:2011mk}.  After the subtraction of Standard Model backgrounds, this excess takes the form of a peak-like feature at approximately 140-150~GeV in the invariant mass distribution of the two jets. We show in Sec.~\ref{sec:Wjj} that this observed peak could be produced by a $Z'$ that has a mass in this energy range and with modest couplings to light quarks ($g_{qqZ'} \sim 0.1$--$0.3$).

In Secs.~\ref{sec:bjets} and~\ref{sec:ttbar} we discuss two other anomalous results from the Tevatron experiments that could also be the result of a new leptophobic $Z'$ in a similar mass range. First, there is a modest excess in the distribution of events with at least three $b$-jets reported by both CDF and D0. While the results of this search channel are usually interpreted in terms of non-Standard Model Higgs phenomenology ({\it i.e.}\ two Higgs doublet models with large $\tan \beta$), we show that the reported excess could also arise from a $Z'$ with a mass between approximately 130 and 160 GeV, and with relatively large couplings to $b$-quarks ($g_{bbZ'} \sim 0.7$--$0.9$). Secondly, CDF has also reported an excess in the $t\bar{t}$ forward-backward asymmetry, inconsistent with the Standard Model at the 3.4$\sigma$ level for $t\bar{t}$ invariant masses above $450$~GeV~\cite{Aaltonen:2011kc} (this excess has recently been confirmed in Ref.~\cite{cdfdilepton}). D0 also finds an asymmetry in tension with the Standard Model, although with less statistical significance~\cite{:2007qb}. While the $s$-channel exchange of a $Z'$ cannot explain the asymmetry (via the process $q\bar{q}\to (Z')^* \to t\bar{t}$) without running afoul of measurements of the total top pair production cross section~\cite{Aaltonen:2009iz,Aaltonen:2010ic,Moch:2008ai}, a flavor violating coupling allowing the $Z'$ to produce $t\bar{t}$ through its $t$-channel exchange could yield the observed asymmetry~\cite{Jung:2009jz,Barger:2010mw,Xiao:2010hm}.

In addition to being interesting new physics in its own right, the existence of a $Z'$ boson could also have important implications for other fields, including cosmology. In Sec.~\ref{sec:DM}, we discuss the role that a $Z'$ could play in dark matter phenomenology. In particular, a relatively light $Z'$ which couples to both dark matter and quarks could lead to large elastic scattering cross sections between dark matter and nuclei, potentially generating high rates at direct detection experiments (such as the those implied by the results reported by the CoGeNT \cite{Aalseth:2010vx} and DAMA~\cite{Bernabei:2010mq} collaborations). The $s$-channel exchange of a $Z'$ could also provide an important dark matter annihilation channel. When the cross section for this process is used to calculate the thermal relic abundance, we find that the predicted contribution from $Z'$ exchange can potentially lead to an abundance of dark matter consistent with the measured cosmological density~\cite{Buckley:2010ve}.

In this paper, we take a model independent approach to the possibility of relatively light and leptophobic $Z'$ gauge bosons. That is to say, we do not assume any specific overall symmetry that would fix the relative couplings of the quarks and leptons. We find that any or all or of the aforementioned Tevatron anomalies could potentially be accounted for by the introduction of a $Z'$ boson. 

%
%

\section{Summary of Current Experimental Constraints \label{sec:bounds}}

Experimental constraints on $Z'$ models are often presented under one of two assumptions: either that the couplings of the $Z'$ are identical to those of the Standard Model $Z$ boson (sometimes scaled by an overall factor), or that the couplings are set within the context of a specific top-down model ($E_6$ or $SO(10)$ GUT models, for example). Here, we take a more agnostic view of the structure of the couplings and treat each of the $Z'$-fermion-fermion couplings as a free parameter. As we will demonstrate, this opens a considerable range of $Z'$ masses and couplings that would be strongly excluded under common assumptions. We begin with a review of the current experimental bounds on the properties of the $Z'$.

Among the strongest constraints on $Z'$ couplings to leptons are those coming from LEP~II. In particular, the process $e^+ e^- \rightarrow Z' \rightarrow e^+ e^-$ leads to a constraint of $g_{eeZ'} \lesssim 0.044 \times (m_{Z'}/200 \, {\rm GeV})$ for $Z'$ masses above roughly 200~GeV~\cite{lep:2003ih,Carena:2004xs,Bando:1987br}. At lower masses, the LEP II constraint, which is derived in an effective field theory formalism, is not directly applicable. Below approximately 200~GeV, off-shell $Z'$ production is no longer suppressed by the $Z'$ mass, but rather by the LEP center-of-mass energy. A conservative constraint is therefore $g_{eeZ'} \lesssim 0.04$ for $m_{Z'} \lsim 200$~GeV. Much stronger constraints can be placed on the production and decay into $e^+ e^-$ pairs of on-shell $Z'$ bosons if the $Z'$ mass is near one of the center-of-mass energies at which LEP~II operated: 130, 136, 161, 172, 183, 189, and 192--209 GeV~\cite{lep:2003ih}. Constraints from the $s$-channel production of $e^+ e^-$~\cite{Abulencia:2006xm} and/or $\mu^+ \mu^-$~\cite{Aaltonen:2008ah} at the Tevatron are also quite stringent ($\tau^+ \tau^-$ final states are considerably less constrained~\cite{Acosta:2005ij}). A $Z'$ with Standard Model-like couplings, for example, must be heavier than approximately 1~TeV to be consistent with the null results of these searches~\cite{Nakamura:2010zzi}.

A $Z'$ that does not couple to leptons as strongly as the Standard Model $Z$ (a leptophobic $Z'$), however, is much more difficult to observe or constrain at both lepton and hadron colliders. Although one would naively expect that a search for a peak in the dijet invariant mass distributions would suffice at a hadron collider, the QCD background at low dijet mass (compared to the beam energy) introduces large theoretical uncertainties, swamping any resonance signal arising from a $Z'$ with electroweak-strength or smaller couplings. For a leptophobic $Z'$ with a mass between $\sim 300-900$~GeV, dijet searches at the Tevatron ($p \bar{p} \rightarrow Z' \rightarrow q \bar{q}$) constrain its couplings to quarks to be comparable to or less than those of the Standard Model $Z$ \cite{Aaltonen:2008dn}. For a leptophobic $Z'$ below 300 GeV, the uncertainties in the QCD background overwhelm the signal at the Tevatron, and so the strongest constraints come from the lower energy UA2 experiment~\cite{Alitti:1993pn}. From the lack of an observed dijet resonance,  UA2 can place constraints on the order of $g_{qqZ'} \lesssim 0.2$--$0.5$ for $Z'$ masses in the range of 130 to 300 GeV.

\begin{figure}[htp]
\centering
\includegraphics[width=0.5\columnwidth]{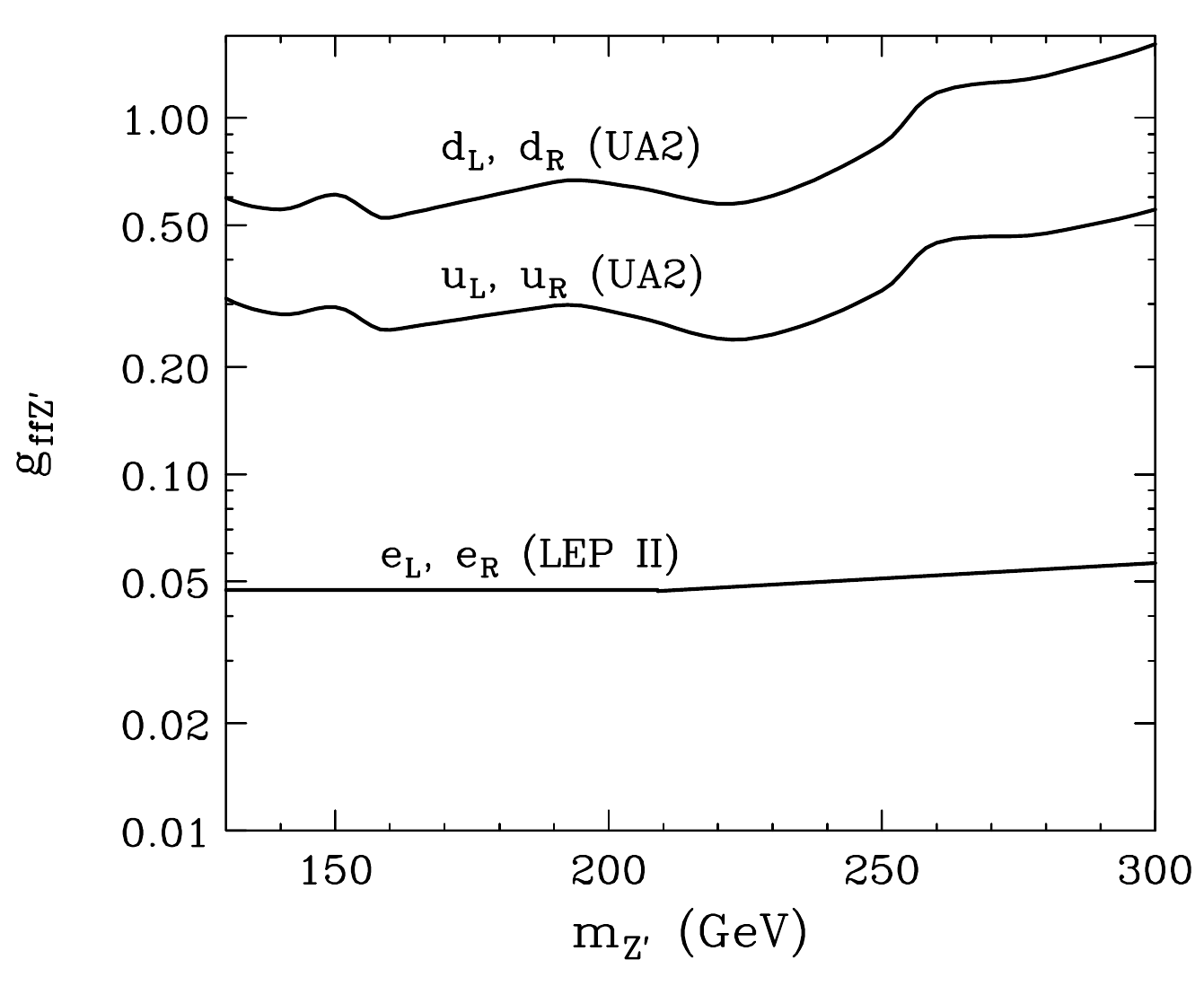}
\caption{Constraints on the $Z'$ couplings to light quarks and leptons as a function of the $Z'$ mass. Bounds on $Z'$ couplings to light quarks were extracted from the results of the UA2 collaboration~\cite{Alitti:1993pn}, whereas the LEP~II bounds on couplings to electrons were derived from Ref.~\cite{Carena:2004xs,Bando:1987br}. We have assumed coupling to a single fermion helicity. The constraints on the couplings of a $Z'$ to leptons are significantly more stringent than those on couplings to quarks.\label{fig:UA2}}
\end{figure}

In Fig.~\ref{fig:UA2}, we show the constraints from UA2 and LEP~II on the couplings of a relatively light $Z'$ to first generation quarks and electrons, assuming couplings to a single helicity. To obtain the UA2 limits, we have computed the cross section for the process $p\bar{p} \to Z' \to 2\ \text{jets}$ at a center of mass energy of 630~GeV using MadGraph/MadEvent~\cite{Alwall:2007st}, and have compared the result to the limits on dijet production shown in Fig.~2 of Ref.~\cite{Alitti:1993pn}. We see from Fig.~\ref{fig:UA2} that a 130--300~GeV $Z'$ with roughly equal couplings to quarks and leptons is constrained by LEP~II to have very small overall gauge coupling and thus will be unlikely to provide any observable signals at the Tevaton, and possibly even the LHC. Phenomenologically much more interesting is the scenario in which a relatively light $Z'$ has very small couplings to electrons and muons ($\lsim 0.04$), but sizable ($\sim 0.1$--$0.3$) couplings to quarks. We will focus on this case throughout the remainder of this paper.

There are also a number of indirect and low energy constraints that restrict the mass and couplings of $Z'$ bosons. In particular, mixing between the $Z'$ and the Standard Model $Z$, which is expected in a wide range of $U(1)'$ models, can shift the $Z$ mass from its predicted Standard Model value, contributing to the $T$ parameter~\cite{Peskin:1991sw} (although the $S, T, U$ parametrization must be used carefully within the context of $Z'$ models, as the electroweak corrections are not generally oblique).  High precision determinations of the $Z$ mass and other electroweak measurements thus strongly constrain the degree of mixing that is allowed between the $Z$ and a light $Z'$~\cite{Langacker:1991pg,Umeda:1998nq}. However, the degree of $Z$--$Z'$ mixing expected is highly model dependent, and there is no \emph{a priori} reason to expect a large mixing angle. To avoid conflict with electroweak precision data, we will assume negligible $Z$--$Z'$ mixing throughout this paper.

If the couplings between the $Z'$ and Standard Model quarks are not family universal, tree-level flavor-changing neutral current processes will be generated~\cite{Langacker:2000ju}.  Measurements of neutral $K$, $D$, and $B$ meson mixing restrict couplings among the first two generations and the $b$ quark to be quite small~\cite{Langacker:2000ju, He:2004it, Gupta:2010wt}.  However, flavor-changing processes involving the top quark are relatively unconstrained by experiment, so that couplings such as $\bar{u}tZ'$ may be substantial.  We will consider this possibility and its implications further in Section~\ref{sec:ttbar}, within the context of the $t\bar{t}$ forward-backward asymmetry measured at the Tevatron.

\section{$W^{\pm} +$Dijet Events at the Tevatron \label{sec:Wjj}}

\begin{figure}[ht]
\includegraphics[width=0.29\columnwidth]{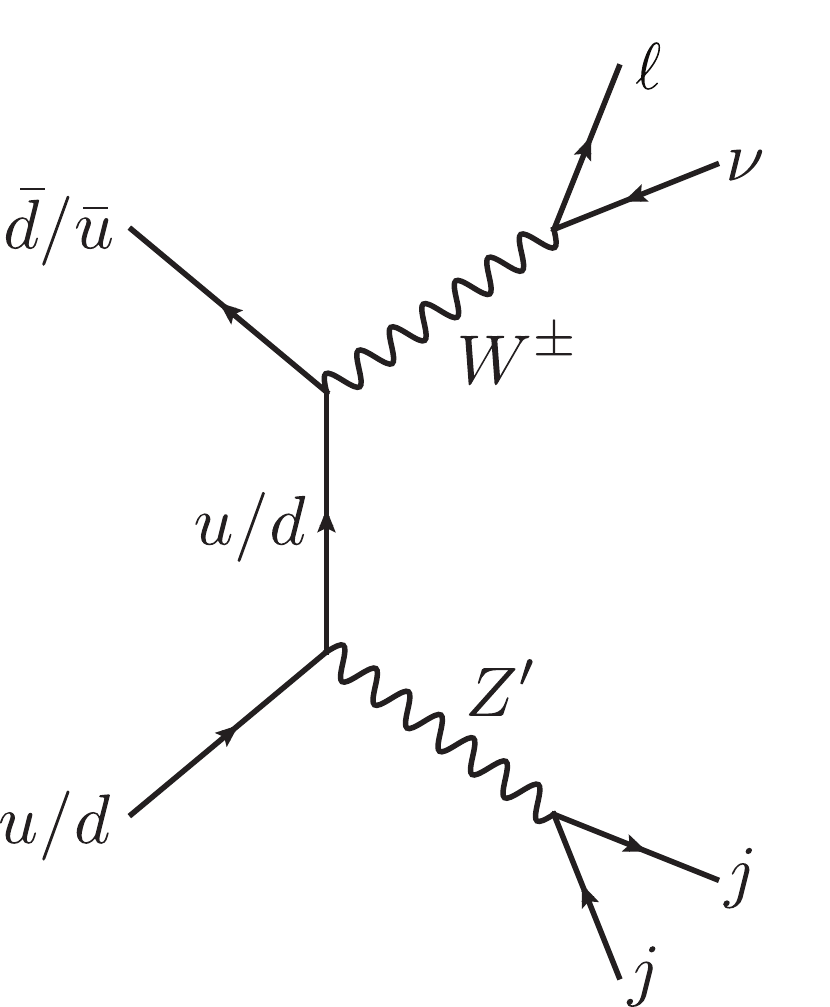}
\caption{A representative Feynman diagram contributing to events containing a lepton, missing transverse energy, and two jets. When plotted as a function of the invariant dijet mass, this process will produce a peak at the mass of the $Z'$. \label{fig:dijetfeyn}}
\end{figure}

The CDF collaboration has recently presented the results of an analysis studying events with a lepton, missing transverse energy, and a pair of hadronic jets \cite{Aaltonen:2011mk}. In the Standard Model such events arise predominantly from QCD processes in which an additional $W^\pm$ decaying to $l^{+} \nu$ or $l^{-}\bar{\nu}$ is radiated. A smaller contribution is due to the production of a $W^\pm$ plus an additional weak gauge boson (another $W^\mp$ or a $Z$) decaying hadronically. When the number of $W^{\pm}\rightarrow  l \nu$ plus two jet events is plotted as a function of the invariant mass of the two jets, $m_{jj}$, a broad peak is found at the masses of the $W^{\pm}$ and the $Z$. The existence of a $Z'$ with significant couplings to Standard Model quarks could lead to the appearance of an additional peak at the mass of the new boson, through processes such as those shown in Fig.~\ref{fig:dijetfeyn}. 

The recent CDF analysis \cite{Aaltonen:2011mk}, which makes use of 4.3~fb$^{-1}$ of data, reports the presence of a feature consistent with such a peak, consisting of 253 events ($156\pm42$ in the electron sample and $97\pm38$ in the muon sample) above expected backgrounds in the sum of the electron and muon channels. The center of the peak is located at a dijet invariant mass of $144\pm 5$~GeV. 
%
%
Relative to searches for dijets resulting from $s$-channel $Z'$ exchange, the requirement of an associated lepton and missing energy (assumed to come from a decaying $W^\pm$) drastically reduces the background. Indeed, this channel is exactly where one would expect to see the first indications of a relatively light leptophobic $Z'$. 

To examine whether the observed excess can be explained by a $Z'$ boson, we have performed simulations using MadGraph/MadEvent, together with Pythia 6~\cite{Sjostrand:2006za}, for parton showering and hadronization, and Delphes~\cite{Ovyn:2009tx} as a detector simulation. The kinematic cuts described in~\cite{Aaltonen:2011mk} are applied. For simplicity, we use a generic set of input parameters for Delphes. That is, we did not implement the actual detector parameters of the CDF experiment, as we find that using the generic parameters already provides a description of the diboson background that is acceptable for the purposes of this study, implying that the detector efficiency and energy resolution are adequately modeled.  

\begin{figure}[htpb]
  \begin{center}
    \includegraphics[width=0.45\columnwidth]{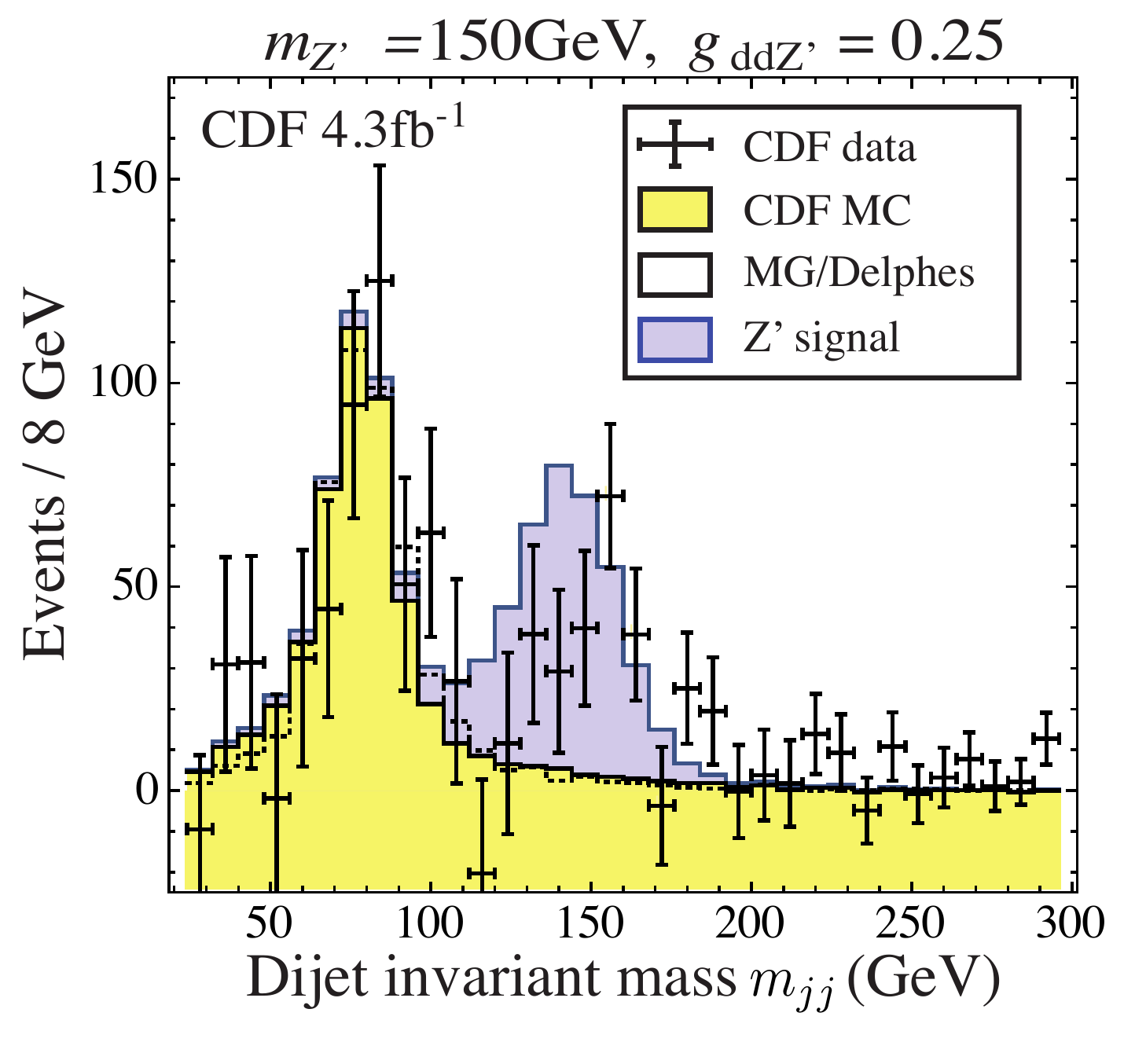}
  \end{center}
  \caption{The observed distribution of $W+\text{dijet}$ events at CDF~\cite{Aaltonen:2011mk} (black data points) after subtraction of all Standard Model backgrounds except those from diboson ($W^+W^-$ or $W^{\pm}Z$) production. Note the good agreement between our prediction of the remaining background (dashed histogram) with the data and with the prediction from the full CDF Monte Carlo simulation (yellow/light gray shaded histogram), which provides some level of confidence in our modeling of detector effects and analysis cuts. The blue/dark gray shaded histogram corresponds to the CDF background prediction, plus a signal from a 150~GeV $Z'$ boson with $g_{ddZ'}=0.25$, $g_{uuZ'}=0$.}
  \label{fig:dijet-spectrum}
\end{figure}

We find that the observed excess of events can be explained by a $Z'$ boson with a mass of $\sim 150$~GeV and with coupling $g_{ddZ'} \sim 0.25$ (for $g_{uuZ'}=0$) or $g_{uuZ'} \sim 0.25$ (for $g_{ddZ'}=0$), leading to a cross section $\sigma (p\bar{p}   \rightarrow Z'+W^{\pm}) \approx 1.8$~pb. This is illustrated in Fig.~\ref{fig:dijet-spectrum}, where we compare the prediction of such a $Z'$ model to CDF data. Note that for the events considered here, only couplings to left-handed quarks are relevant due to the presence of a $W^\pm$.

\begin{figure}[ht]
  \begin{center}
    \includegraphics[width=0.4\columnwidth]{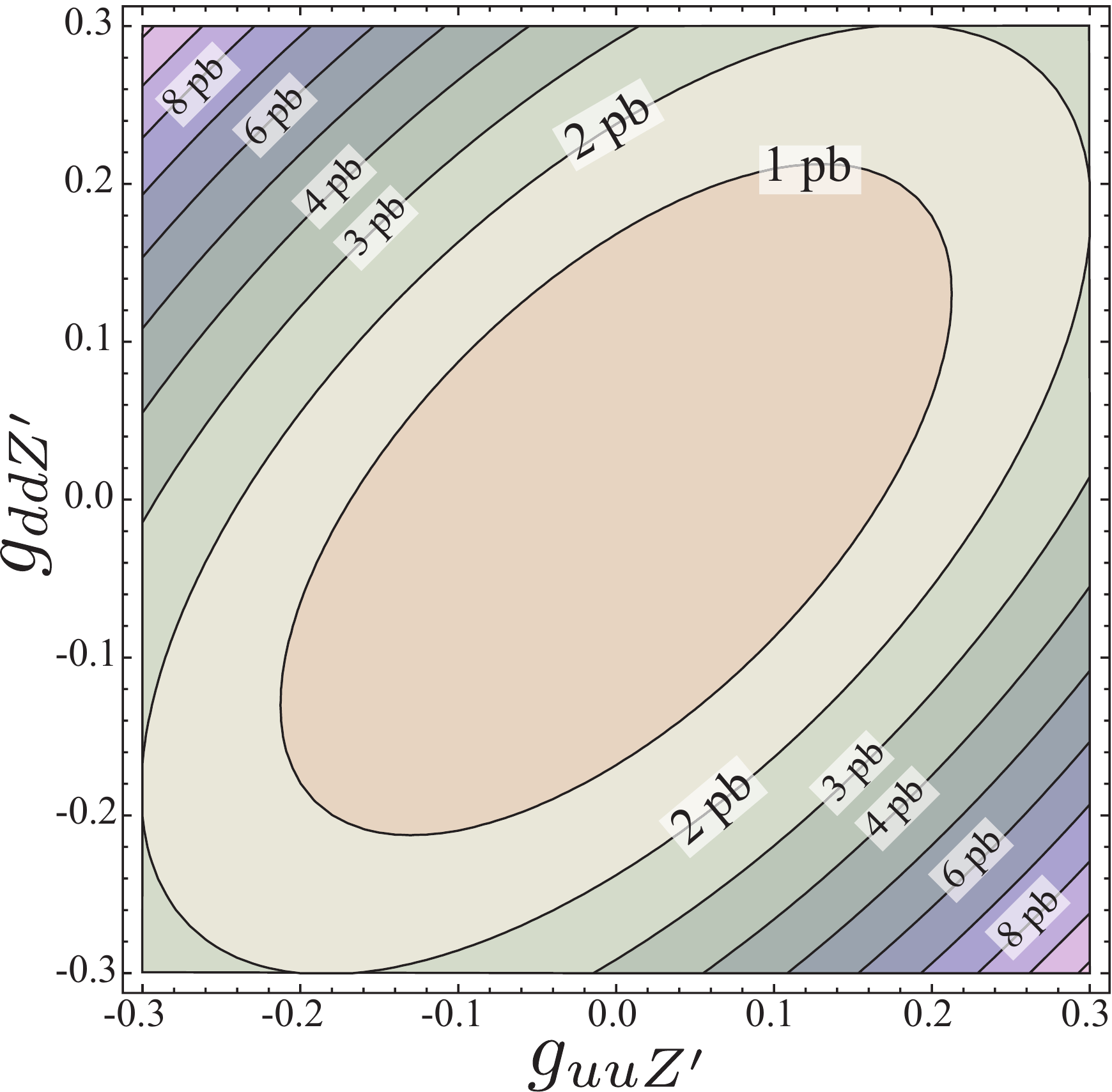}
  \end{center}
  \caption{Contour plot of the production cross section $\sigma (p\bar{p} \rightarrow Z'+W^{\pm})$ as a function of the couplings of the $Z'$ to left-handed first generation quarks, $g_{uuZ'}$ and $g_{ddZ'}$, for a mass of $m_{Z'}=150$ GeV. Note that interference effects occur when both couplings are non-zero.\label{fig:dijetsigma}}
\end{figure}

The $W^{\pm}+\text{dijet}$ cross-section (before cuts) as a function of $g_{uuZ'}$ and $g_{ddZ'}$ is shown in Fig.~\ref{fig:dijetsigma} (as computed using FeynArts and FormCalc~\cite{Hahn:2000kx}).  It should be noted that the cross-section is actually reduced if $g_{uuZ'}$ and $g_{ddZ'}$ are equal, due to the presence of interference terms between the two diagrams with $\bar{u} d$ and $\bar{d} u$ initial states (see Fig.~\ref{fig:dijetfeyn}).  On the other hand, if the two couplings are taken to have opposite sign, then the interference enhances the $W^{\pm}+\text{dijet}$ cross-section.  The value  $\sigma (p\bar{p}   \rightarrow Z'+W^{\pm}) \approx 1.8$~pb leading to the results shown in Fig.~\ref{fig:dijet-spectrum} can be obtained with e.g. $g_{uuZ'} = - g_{ddZ'} \sim 0.13$.

We note that evidence of such a $Z'$ could also come from other channels including two jets plus missing energy, two jets plus a photon, or two jets plus two leptons. At the current level of precision, these channels do not yet impose a strong constraint, but in the future could provide interesting avenues for testing leptophobic $Z'$ models.  

\section{Multi-$b$ Events at the Tevatron \label{sec:bjets}}

\begin{figure}[ht]
 \includegraphics[width=0.275\columnwidth]{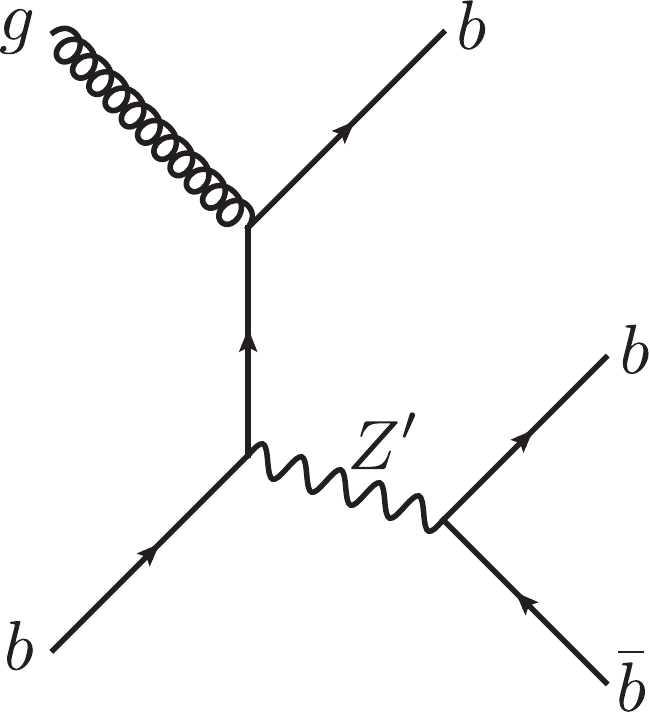}


\caption{One of the dominant diagrams leading to final states with three $b$-jets through an on-shell $Z'$. Four-$b$ final states (with at least three $b$-tags) can also contribute to the analysis discussed here, but are subdominant in the parameter range of interest. \label{fig:threebfeyn}}
\end{figure}

Feynman diagrams similar to those leading to the production of $W^{\pm} Z'$ at the Tevatron could also provide potentially observable signals in other channels. In particular, if we allow the $Z'$ to have large couplings to $b$ quarks, collisions at the Tevatron can lead to events with three $b$-jets, through diagrams such as that shown in Fig.~\ref{fig:threebfeyn}. In such a scenario, a $Z'$ could be observed in searches designed to look for Higgs bosons decaying to $b$-quark pairs.

The CDF and D0 collaborations have each presented results from their searches for Higgs bosons decaying to $b\bar{b}$ in association with an additional $b$-jet. Based on 2.2 fb$^{-1}$ of data, the CDF collaboration has reported a limit that is more than 2$\sigma$ weaker than expected for Higgs masses in the range of 130 to 160 GeV (with an estimated probability of obtaining such a weak limit for any mass estimated to be 5.7\%)~\cite{cdf3b}. The D0 analysis, based on 5.2 fb$^{-1}$ of data, observes a small (and not particularly statistically significant) excess in their 3-$b$ channel at Higgs masses of roughly 100-250 GeV~\cite{Abazov:2010ci}. To attribute either or both of these excesses to a Higgs boson is quite difficult: requiring a cross section that is much larger than is predicted in the Standard Model, or even within two Higgs doublet models with large $\tan \beta$. In this section, we will discuss the possibility that a $Z'$ with the properties needed to produce the $W^{\pm}+\text{dijet}$ signal observed by CDF could also be responsible for these excesses of multi-$b$ events. \footnote{An alternative model which can provide a joint explanation of these two anomalies through the addition of heavy color-octet particles is given by \cite{Bai:2010dj}.}

From Refs.~\cite{cdf3b,Abazov:2010ci}, we estimate that $\sigma(p\bar{p} \rightarrow H+b)\times {\rm BR}(H\rightarrow b\bar{b}) \sim 5$--$10$~pb is required to yield a signal capable of reconciling the data with the theoretical prediction. If interpreted as events involving a $Z'$ rather than a Higgs boson, the impact of kinematic cuts (and $b$-tagging efficiencies) is modified, and thus the underlying cross section times branching ratio is affected. Using Madgraph/MadEvent combined with Pythia 6 to account for the kinematic cuts described in the analysis, we find that the cross section times branching ratio, $\sigma(p \bar{p} \rightarrow Z' + b) \times \, {\rm BR}(Z' \rightarrow b\bar{b})$, needed to account for these events is approximately 25\% smaller than that required of Higgs associated events. If the couplings of the $Z'$ with all species of quarks is set to $g_{qqZ'} \approx 0.2$ (the approximate values needed to generate the dijet excess at CDF), we calculate $\sigma(p \bar{p} \rightarrow Z' + b \bar{b})\sim 1$~pb, which is well below what is needed to produce the multi-$b$ excess. To increase the number of multi-$b$ events to the desired value requires considerably larger couplings between the $Z'$ and $b$ quarks. In particular, we find that a value of $g_{bbZ'} \sim 0.7-0.9$ (to either left- or right-handed $b$'s) is required to produce the observed excesses. 


As a general point, $Z'$ couplings that are large enough to produce the observed multi-$b$ signals invariably lead to large branching ratios of the $Z'$ into $b\bar{b}$. For example, the particular choice of left-handed couplings $g_{bbZ'} = 0.7$, $g_{uuZ'}=g_{ddZ'} = 0.2$ leads to a $Z'$ branching ratio to $b$'s of $\sim 75\%$. The resulting high multiplicity of $b$'s in the $Z'$ decay should also be observable in the CDF dijet analysis.  While this paper was in preparation, CDF has reported that the fraction of $W$ + dijet events with the dijets identified as $b$ quarks is not significantly higher in the excess region compared to the sideband regions \cite{Aaltonen:2011mk}.  Further studies from CDF and D0 will be needed in order to assess whether using a single $Z'$ to explain both the $W + jj$ and multi-$b$ anomalies is still feasible. Due to the lower backgrounds of events that include $b$-jets, searches in channels such as two $b$-jets plus missing energy, two $b$-jets plus a photon, or two $b$-jets plus two leptons, could also be fruitful in identifying a $Z'$ with sizable couplings to $b$-quarks.

The presence of a $Z'$ with relatively large coupling to $b$ quarks can lead to unwanted shifts in precision electroweak quantities~\cite{:2005ema}, in particular the branching ratio of $Z \rightarrow b\bar{b}$.  For a $Z'$ coupling only to left-handed $b$ quarks, the loop contribution to this branching ratio is approximately equal to the current experimental error bar for $g_{bbZ'} \sim 0.7$.  On the other hand, if the $Z'$ coupling is taken to be right-handed, the loop contribution is suppressed by a factor of $\sim 30$ due to the smaller coupling of the Standard Model $Z$ to right-handed quarks. In this case, there is effectively no constraint on the $Z'$ coupling strength. Similar loop contributions to other quantities are less constraining.  In particular the $Z'$ loop contribution to the bottom quark forward-backward asymmetry, which remains in slight tension with global precision electroweak fits, is much smaller than the current experimental error.

\section{$t-\bar{t}$ Forward-Backward Asymmetry At The Tevatron \label{sec:ttbar}}

The forward-backward asymmetry in top quark pair production at the Tevatron has first been studied by the D0 and CDF experiments in Refs.~\cite{Abazov:2007qb,Aaltonen:2008hc}, and recently measured by CDF using a significantly larger dataset~\cite{Aaltonen:2011kc}. This new analysis finds a $3.4\sigma$ discrepancy between the prediction of the Standard Model and the asymmetry measured in events with a large $t\bar{t}$ invariant mass (the discrepancy is less than $2\sigma$ if all values of the $t\bar{t}$ invariant mass are included). A more recent analysis by CDF identifies further evidence for such a discrepancy among $t\bar{t}$ dilepton events~\cite{cdfdilepton}.

\begin{figure}
  \begin{center}
  \begin{tabular}{m{4cm}m{2cm}m{4cm}}
\includegraphics[width=0.2\columnwidth]{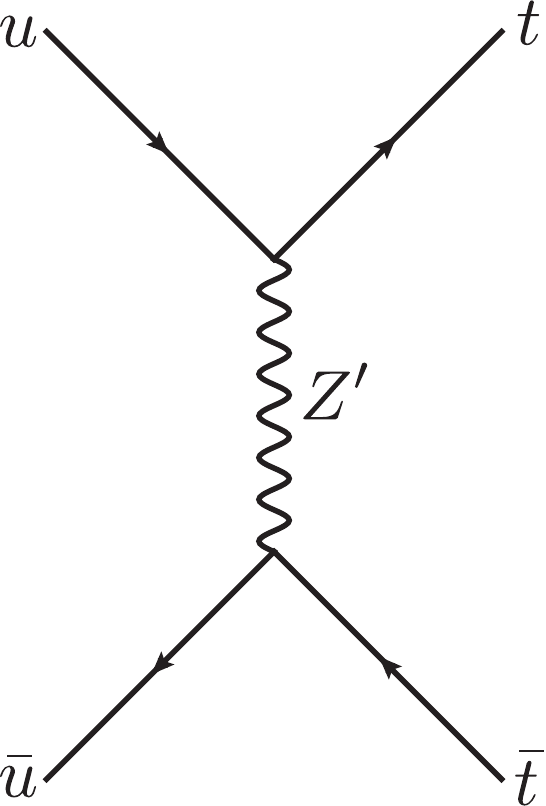}&&\includegraphics[width=0.2\columnwidth]{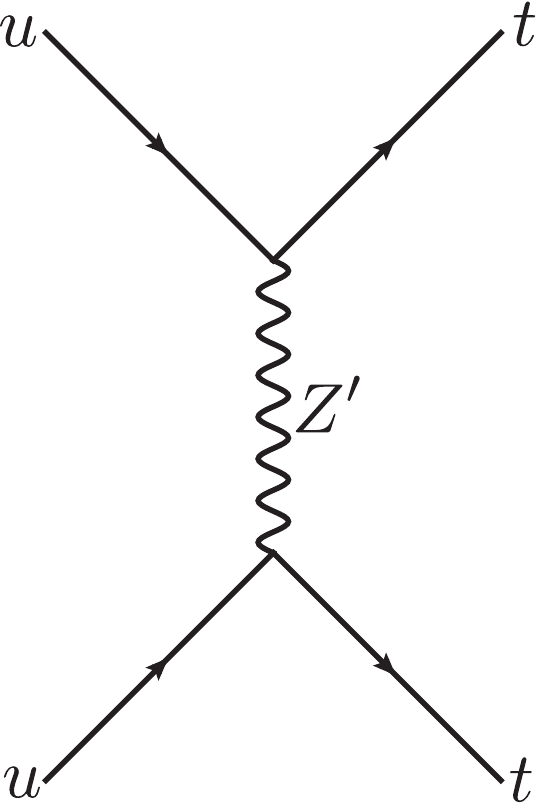}\\
\end{tabular}

\caption{{Left panel: this diagram contributes to the $t\bar{t}$ asymmetry, through interference with gluonic contributions to the same process in the Standard Model.  Right panel: a similar diagram gives rise to same-sign top quark production at tree level.  Experimental constraints on $tt, \bar{t}\bar{t}$ production must be taken into account when attempting to explain the observed $t\bar{t}$ asymmetry.} \label{fig:ttfeyn}}
  \end{center}
\end{figure}

One possible explanation for this discrepancy is a flavor-violating chirally
coupled $Z'$ boson that mixes, for instance, up and top quarks~\cite{Jung:2009jz,
Cao:2010zb, Cao:2011ew, Berger:2011ua, Bhattacherjee:2011nr, Barger:2011ih,
Gresham:2011dg}.\footnote{An alternative possibility is a $W'$ boson coupling
down and top quarks~\cite{Cheung:2009ch,Cheung:2011qa}. Alternatively, axigluons~\cite{Antunano:2007da,Frampton:2009rk,Isidori:2011dp,
Bai:2011ed} or other heavy color multiplets~\cite{Arhrib:2009hu,Shu:2009xf,Dorsner:2009mq,Grinstein:2011yv, Patel:2011eh} (but see also \cite{Ligeti:2011vt}) 
can be invoked to explain the CDF $t\bar{t}$ asymmetry.}
Since such a $Z'$ boson contributes to $t\bar{t}$ production only
in the $t$-channel (as shown in Fig.~\ref{fig:ttfeyn}), it will not necessarily lead to unacceptable modifications to the total $t\bar{t}$ cross section, although contributions to other processes such as same-sign top production (also shown in Fig.~\ref{fig:ttfeyn}) must be taken into account.

\begin{figure}
  \begin{center}
    \includegraphics[width=0.42\columnwidth]{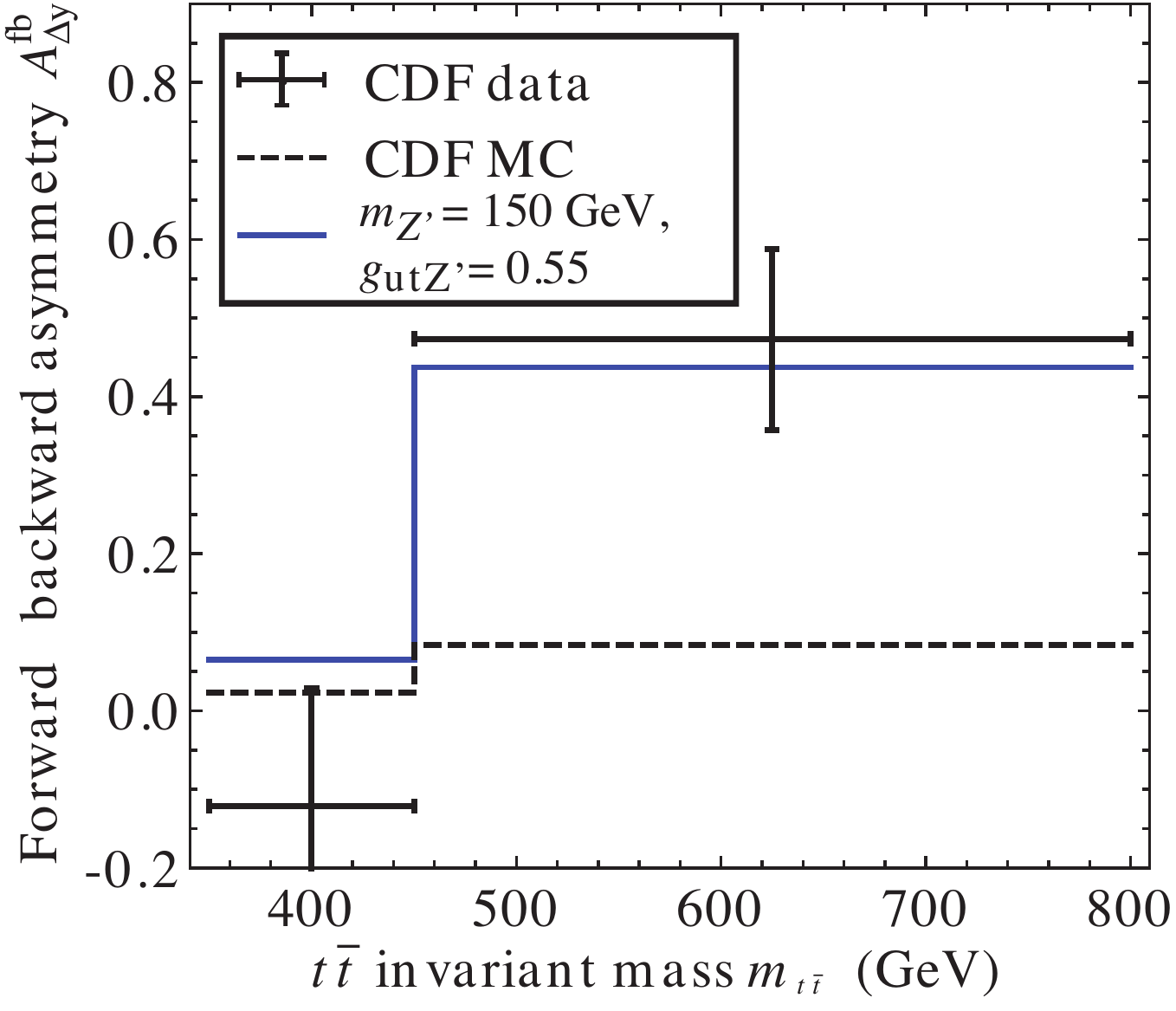}\includegraphics[width=0.42\columnwidth]{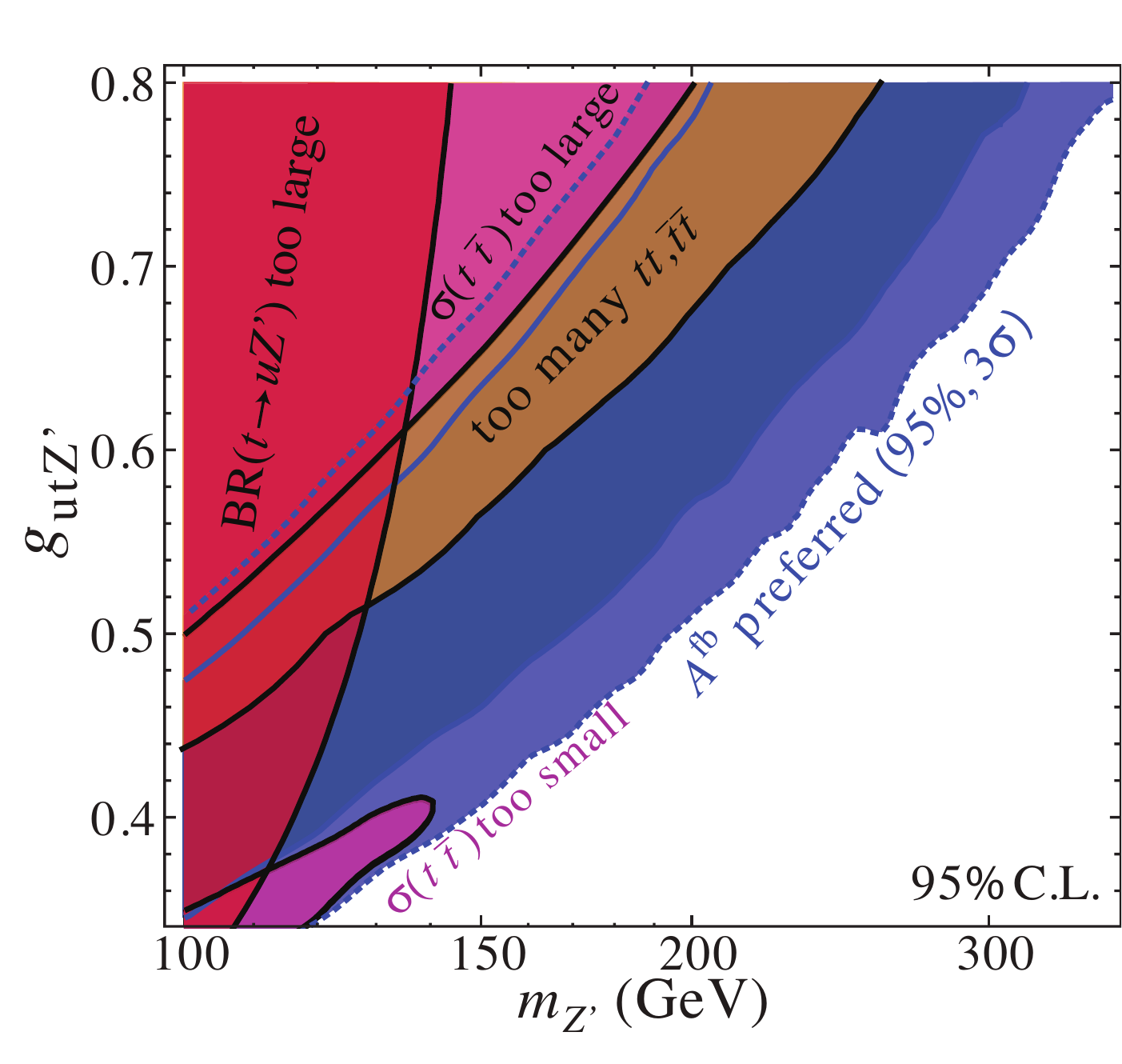}
  \end{center}
  \caption{Left panel: Top-antitop asymmetry $A^{t\bar{t}}$ predicted by the
    Standard Model (black dotted histogram), by the Standard Model extended by
    a representative $Z'$ boson with couplings according to
    Eq.~\eqref{eq:AttZprimeOp} (blue solid histogram), and CDF measurement of
    $A^{t\bar{t}}$ (data points with error bars). Right panel: Favored
    parameter regions in the $Z'$ model according to
    Eq.~\eqref{eq:AttZprimeOp}. Contours are computed using the $\Delta \chi^2$
    method with 2 degrees of freedom. We also show constraints from same-sign top events, from
    the total $t\bar{t}$ production cross section, and from the non-standard
    top decay mode $t \to u Z'$ (assuming the $Z'$ decays into jets).  Our best fit regions are in good agreement with the results of Ref.~\cite{Jung:2009jz}.}
  \label{fig:mttAtt}
\end{figure}

To assess in more detail the consistency of a flavor-violating $Z'$ boson with
the CDF data, we consider the model proposed in Ref.~\cite{Jung:2009jz} in which the $Z'$ couples through the
operator
\begin{align}
  g_{utZ'}  Z_\mu' \bar{u} \gamma^\mu P_R t  +  h.c. \,,
  \label{eq:AttZprimeOp}
\end{align}
where $P_R = (1 + \gamma^5)/2$ is the projector onto right-chiral states, and
$g_{utZ'}$ is the flavor-violating $Z'$ coupling constant. We have simulated
tree-level $t\bar{t}$ production in this model at the parton-level using
MadGraph/MadEvent. We compute the $t\bar{t}$
asymmetry in the $t\bar{t}$ rest frame as
\begin{align}
  A^{t\bar{t}}_{Z'} = \frac{N(\Delta y > 0) - N(\Delta y < 0)}{N(\Delta y > 0) + N(\Delta y < 0)} \,,
  \label{eq:Att}
\end{align}
where $N(\Delta y \lessgtr 0)$ is the number of events in which the rapidity
difference between the top and the anti-top quark is less/greater than zero.
Since our simulation is carried out at tree level, it includes only the new physics
contribution to the asymmetry, but not the Standard Model terms which arise
at next-to-leading order. To compare our predictions to CDF data, we
therefore add the Standard Model asymmetry, which we take from
Ref.~\cite{Aaltonen:2011kc}.  In the left panel of Fig.~\ref{fig:mttAtt}, we
show the $t\bar{t}$ asymmetry predicted in the $Z'$ model (including the Standard Model contribution) for $m_{Z'} = 150$~GeV, $g_{utZ'} = 0.5$, and compare it to
CDF data and to the Standard Model prediction alone. We observe that the $Z'$
model can explain the increase of the asymmetry with increasing $t\bar{t}$
invariant mass $m_{t\bar{t}}$.

In the right panel of Fig.~\ref{fig:mttAtt}, we show the preferred
regions of the $Z'$ parameter space. We find that a $Z'$ with a mass between 100 and
300~GeV, and couplings $g_{utZ'}$ on the order of $\sim 0.3$--$0.8$ provides the
best fit to the experimental observations. We also show the constraints
on the $Z'$ model coming from measurements of the total cross section of
$t\bar{t}$ production, $\sigma(t\bar{t})$, same-sign top
production, and the non-standard top
decay mode $t \to u Z'$. If Eq.~\eqref{eq:AttZprimeOp} is the only operator
coupling the $Z'$ to the Standard model, the $Z'$ will be long lived, and will lead to a
large amount of missing energy in top decays involving a $Z'$. Here, however, we assume that it has
additional couplings to light quarks or $b$'s, as in Secs.~\ref{sec:Wjj} and \ref{sec:bjets}.

To derive the total cross section $\sigma(t\bar{t})$, we
use the approximate next-to-next-to-leading order Standard Model prediction of
$\sigma(t\bar{t})$~\cite{Moch:2008ai} and add to it the difference
between the tree-level cross section with and without
inclusion of the $Z'$. We then compare this number to the combined result of
several CDF measurements~\cite{Nakamura:2010zzi}, taking into account both the
experimental error and the theoretical uncertainty from Ref.~\cite{Moch:2008ai}.  
Note that, in the preferred region of the
$m_{Z'}$--$g_{utZ'}$ parameter space, the $Z'$ model actually predicts a
\emph{decrease} in $\sigma(t\bar{t})$, due to interference effects~\cite{Jung:2009jz}.
In fact this can lead to an excluded region at low $Z'$ mass and small coupling
where the total cross section is too small. It has recently been pointed out, however, that the selection efficiency of $t\bar{t}$ events may differ in various models, potentially altering these constraints somewhat~\cite{Gresham:2011pa}.

For the constraint from same-sign top pairs, we compare the cross section for
the processes $p\bar{p} \to tt, \bar{t}\bar{t}$ predicted by
MadGraph/MadEvent to the experimental constraint from
CDF~\cite{Aaltonen:2008hx}, taking into account the 0.5\% acceptance of the
experimental analysis, and using the Feldman-Cousins
method~\cite{Feldman:1997qc,Giunti:1998xv} for the statistical analysis.  The resulting
95\% C.L. exclusion contour is shown in Fig.~\ref{fig:mttAtt}.  The constraint shown does not take into account the process $uu \rightarrow Z'Z'$, which can then lead to same-sign top production if $m_{Z'} > m_t$ and the branching ratio BR$(Z' \rightarrow \bar{u} t)$ is substantial
(we assume this branching ratio to be small).

While this work was being completed, the CMS collaboration
has announced results from a same-sign dilepton
search with 35 pb${}^{-1}$ of data~\cite{Chatrchyan:2011wb}.\footnote{We thank Paddy Fox for drawing our
attention to these results.} The flavor-violating Z' model would predict a significant
number of same-sign dileptons from same-sign top production, and is
therefore in severe tension with the null result from CMS. A full
analysis will be required to determine how severe this tension is.

The branching ratio for the top quark decay mode $t \to u Z'$ is only
constrained indirectly.  In particular, CDF has measured the $t\bar{t}$
production cross section independently in events with two charged leptons, jets
and missing energy~\cite{CDF:tll} (interpreted as both top quarks decaying to
$\ell \nu b$) and in events with only one charged lepton, jets and missing
energy~\cite{CDF:tlj} (interpreted as one top decaying to $\ell \nu b$, the
other to jets). If the $Z'$ decays to jets, as assumed here, the decay mode
$t \to u Z' \to 3j$ would contribute to the second of
these measurements, but not to the first one.  We compute the expected ratio of
the two cross sections including the effect of the $Z'$, and compare to the
ratio of the CDF measurements. Note that we do not account for the modified
probability of obtaining a $b$-tag in the $Z'$ model since it depends on
details of the $Z'$ model not relevant to the $t\bar{t}$ asymmetry. If the $Z'$
decays predominantly to light quarks, there will be fewer $b$-tags than in
Standard Model top decays and the bound will be weakened.  If the dominant $Z'$
decay mode is into $b$ quarks, the bound becomes stronger.

An additional constraint comes from the production of single top quarks, which are measured by  both CDF and D0 \cite{Abazov:2009ii, Lueck:2010zz}.  In the presence of a $Z'\bar{u}t$ coupling, single tops can be produced via the processes $p\bar{p} \rightarrow tZ'$ and $p\bar{p} \rightarrow t\bar{u}$.  With a $Z'$ mass and couplings in the range being considered, the production cross-section due to the $Z'$ sector is on the order of the Standard Model prediction $\sigma \sim 3$ fb.  However, as pointed out in Ref.~\cite{Jung:2009jz}, the data analysis in both cases relies upon multivariate analysis techniques optimized for single top production in the Standard Model in order to overcome large backgrounds.  It is therefore difficult to determine whether existing single top observations provide constraints on this $Z'$ model or not.

Finally, the recent CDF analysis \cite{Aaltonen:2011kc} demonstrates a strong dependence of $A^{t\bar{t}}$ on the top/anti-top rapidity difference $\Delta y$.  Specifically, for $|\Delta y| < 1$ they report $A^{t\bar{t}} = 0.026 \pm 0.118$, while for $|\Delta y| \geq 1$ the observed asymmetry is $A^{t\bar{t}} = 0.611 \pm 0.256$.  For the representative point $m_{Z'} = 150$ GeV, $g_{utZ'} = 0.55$, we find
\begin{equation}
\begin{cases}
A_{pred}^{t\bar{t}}(|\Delta y| < 1) &= 0.0505, \\
A_{pred}^{t\bar{t}}(|\Delta y| \geq 1) &= 0.6862, \end{cases}
\end{equation}
where our prediction includes the Standard Model contribution as given in \cite{Aaltonen:2011kc}.  These results are in close agreement with the experimental values.

\section{Implications For Dark Matter \label{sec:DM}}

The existence of a new neutral gauge boson with couplings to both dark matter and the Standard Model could potentially play an important role in dark matter phenomenology. In particular, a $Z'$ could mediate dark matter self-annihilations, as well as the elastic scattering of dark matter with nuclei. In this section, we discuss some of the ways in which a $Z'$ could most significantly contribute to such processes.

If the dark matter particle is either a scalar or a Dirac fermion (which we will denote by the symbol $\chi$), a $Z'$ can mediate a spin-independent elastic scattering cross section with nucleons that is given by
\begin{eqnarray}
\sigma_{p,n} &\approx& \frac{m^2_{\chi} \, m_{p,n}^2 \, g_{\chi\chi Z'}^2}{\pi (m_{\chi}+m_{p,n})^2 M_{Z'}^4} \left[ \left(\begin{array}{c} 2 \\ 1 \end{array}\right) g_{uuZ'}+ \left(\begin{array}{c} 1 \\ 2 \end{array}\right) g_{ddZ'} \right]^2 \nonumber \\
&\approx& 2 \times 10^{-40} \, {\rm cm}^2  \times \bigg(\frac{150\,{\rm GeV}}{m_{Z'}}\bigg)^4\,\bigg(\frac{g_{\chi \chi Z'}}{0.1}\bigg)^2 \, \bigg(\frac{g_{qqZ'}}{0.13}\bigg)^2, 
\end{eqnarray}
where the upper (lower) numbers refer to the cross section with protons (neutrons), and $g_{\chi\chi Z'}$ and $g_{qqZ'}$ denote the $Z'$ couplings to dark matter and light quarks ($u$,$d$), respectively. Thus we see that for couplings needed to produce the observed dijet signal at CDF ($g_{qqZ'} \approx 0.13$) and similar couplings to dark matter ($g_{\chi \chi Z'} \approx 0.1$), we find an elastic scattering cross section similar to that needed to generate the signals reported by the CoGeNT~\cite{Aalseth:2010vx} and DAMA~\cite{Bernabei:2010mq} collaborations (see also Refs.~\cite{cdmslow,Collar:2011kf}). If such a dark matter candidate were heavier than $\sim 8$--$10$~GeV, however, its couplings to the $Z'$ would have to be considerably suppressed in order to evade the constraints from null results of other direct detection experiments~\cite{cdms,xenon}. If the dark matter instead consists of a Majorana fermion, a $Z'$ could also mediate a potentially sizable spin-dependent interaction with nuclei.

%

For dark matter composed of Dirac fermions, the exchange of a $Z'$ can also yield a substantial contribution to its self-annihilation cross section~\cite{Beltran:2008xg,Boehm:2003hm}:
\begin{eqnarray}
\sigma v = \frac{m^2_{\chi} g^2_{\chi\chi Z'}}{2 \pi [(M^2_{Z'}-4 m^2_{\chi})^2]+\Gamma^2_{Z'} M^2_{Z'}} \sum_f g^2_{ffZ'} c_f (1-m^2_f/m^2_{\chi})^{1/2} (2+m^2_f/m^2_{\chi}),
\label{zprimesigmav}
\end{eqnarray}
where $c_f=3$ for quarks and 1 for leptons. For the couplings needed to produce the $W^{\pm} + \text{dijet}$ and multi-$b$ jet signals at the Tevatron ($g_{bbZ'}=0.8$ and $g_{qqZ'}=0.13$, where $q=u,d,s,c$), this leads to an annihilation cross section dominated by $b\bar{b}$ final states, and given by 
\begin{eqnarray}
\sigma v & \approx & 2 \times 10^{-26} \, {\rm cm}^3/{\rm s} \times \bigg(\frac{m_{\chi}}{10\, {\rm GeV}}\bigg)^2 \, \bigg(\frac{g_{\chi\chi Z'}}{0.1}\bigg)^2 \, \bigg(\frac{g_{bbZ'}}{0.8}\bigg)^2 \, \bigg(\frac{150 \, {\rm GeV}}{m_{Z'}}\bigg)^4.
\label{zprimesigmav2}
\end{eqnarray}
This calculation yields a result that is quite similar to the the value required of a simple thermal relic ($3 \times 10^{-26} \, {\rm cm}^3/{\rm s}$). In order for such a dark matter particle to annihilate significantly to $\tau^+ \tau^-$, as would be needed to explain the gamma ray emission observed from the Inner Galaxy~\cite{gc}, another annihilation channel would likely be required. For a dark matter candidate in the form of a scalar, the annihilation cross section is suppressed by the square of the relative velocity. In this case, additional annihilation channels will be necessary to avoid the overproduction of dark matter in the early universe.

\section{Discussion and Conclusions \label{sec:conclusion}}

In this article, we have discussed the phenomenology of a relatively light ($\sim 100$--$200$~GeV) $Z'$ boson, focusing on model independent constraints, and on ways in which such particles could be observed at hadron colliders. Although $Z'$ bosons with sub-TeV masses and Standard Model-like couplings to electrons and muons are excluded by constraints from LEP-II and the Tevatron, we emphasize that much lighter $Z'$ bosons are in fact possible if they couple more weakly to electrons and muons. In particular, we have considered the case of a leptophobic $Z'$ with couplings to electrons and muons that are less than $g_{llZ'} \lsim 0.04$, but with couplings to light quarks that are as large as $g_{qqZ'} \sim 0.25$. Although a light $Z'$ boson with couplings in this range is not currently experimentally excluded, such a particle could potentially be observed in a number of channels at the Tevatron or Large Hadron Collider. 

Within this context, we have discussed three anomalies recently observed at the Tevatron: The 3.2$\sigma$ excess in the distribution of two jet plus $W^{\pm}$ events reported in \cite{Aaltonen:2011mk}, the roughly 2$\sigma$ excess of events with at least three $b$-jets observed by both CDF and D0, and the 3.4$\sigma$ discrepancy between the top quark forward-backward asymmetry measured by CDF and the prediction of the Standard Model. In Table~\ref{tab:summary}, we summarize the mass and couplings of a $Z'$ boson that would be required to account for each of the observed Tevatron anomalies. In the case of the dijet+$W^{\pm}$ excess, the location of the bump-like feature in the distribution of the invariant mass of the jet pairs (see Fig.~\ref{fig:dijet-spectrum}) allows us to constrain the required mass of the $Z'$ to the range of roughly 140 to 150 GeV. To normalize the overall rate of such events, we require the $Z'$ to couple to light quarks with a strength of approximately $g_{qqZ'} \approx 0.1-0.3$ (see Sec.~\ref{sec:Wjj} for details). A $Z'$ of the same mass could also account for the observed excess of multi-$b$ events, but only if it possesses a relatively large coupling to $b\bar{b}$. In this case, we predict that a large fraction of the dijets observed would consist of pairs of $b$-quark jets. Lastly, the forward-backward asymmetry observed in the top quark pair production at the Tevatron (which should shortly be within LHC reach \cite{AguilarSaavedra:2011vw,Hewett:2011wz,Craig:2011an,Bhattacherjee:2011nr,Bai:2011ed}) could also be accounted for with a relatively light $Z'$, although this requires the introduction of a fairly large flavor violating coupling, $g_{utZ'}$.  

\begin{table}
  \centering
  \begin{tabular}{l@{\qquad}c@{\qquad}c@{\qquad}c@{\qquad}c}
    \hline
                         & $M_{Z'}$     & $g_{qqZ'}$ & $g_{bbZ'}$      & $g_{utZ'}$ \\\hline
    $W^{\pm} + jj$             & 140--150~GeV &0.1--0.3 &                  &           \\
    multi-$b$            & 130--160~GeV       & $\ll 1$  & 0.7--0.9       &           \\
    $t\bar{t}$ asymmetry & 120--280~GeV &            &                  & 0.3--0.8  \\\hline
  \end{tabular}
  \caption{Approximate values of the $Z'$ parameters required to explain various excess signals at the Tevatron. Here, quark-quark-$Z'$ couplings refer to a single helicity (left-handed in the case of the $W^{\pm}+jj$ signal, and either left- or right-handed in the other two cases) quarks only.}
  \label{tab:summary}
\end{table}

Intriguingly, we find that each of these anomalies can be simultaneously explained by a $Z'$ with an approximate mass of 140-150 GeV, modest couplings to light quarks, and larger couplings to $b\bar{b}$ and to $u\bar{t}$ and $\bar{t}u$. We emphasize, however, that any subset of these signals could arise from a $Z'$, and that none of these signals need imply the appearance of the others.

If these anomalies at the Tevatron are in fact the result of a $Z'$ with the characteristics described in Table~\ref{tab:summary}, such a particle could have important implications for dark matter phenomenology. In particular, if the dark matter consists of either a scalar or a Dirac fermion with any significant coupling to the $Z'$, it would possess a large elastic scattering cross section with nuclei. With a coupling to the $Z'$ on the order of 0.1, for example, the dark matter would be predicted to possess a cross section with nucleons of $\sigma \sim 2 \times 10^{-40}$ cm$^2$, providing a potential explanation for the signals reported by the CoGeNT and DAMA/LIBRA collaborations. If the dark matter is a Dirac fermion with this same coupling, the large coupling of the $Z'$ to $b\bar{b}$ that is needed to generate the observed multi-$b$ events at CDF would also lead to a dark matter annihilation cross section of $\sigma v \sim 2 \times 10^{-26}$ cm$^3$/s; a value very similar to that needed to thermally produce the measured cosmological dark matter abundance. 

\section*{Acknowledgements} 

We would like to thank Johan Alwall, Paddy Fox, Graham Kribs, Adam Martin, and William Wester for valuable discussions. This work has been supported by the US Department of Energy and by NASA grant NAG5-10842. Fermilab is operated by Fermi Research Alliance, LLC under Contract No.~\protect{DE-AC02-07CH11359} with the US~Department of Energy.

\bibliography{zprime22}
\bibliographystyle{apsrev}

\end{document}